\documentclass[fleqn,usenatbib]{mnras}

\usepackage{newtxtext,newtxmath}

\usepackage[T1]{fontenc}
\usepackage{ae,aecompl}

\usepackage{graphicx}	
\usepackage{amsmath}	
\usepackage{amssymb}	

\title[Modes and oscillations in PSR B1828$-$11]{Mode switching and oscillations in PSR B1828$-$11}

\author[I. H. Stairs et al.]{
I.~H. Stairs,$^{1}$\thanks{E-mail: stairs@astro.ubc.ca (IHS)}
A.~G. Lyne,$^{2}$
M. Kramer,$^{3,2}$
B.~W. Stappers,$^{2}$
\newauthor
J. van Leeuwen,$^{4,5}$
A. Tung,$^{1,6}$
R. N. Manchester,$^{7}$ G. B. Hobbs,$^{7}$
\newauthor
D. R. Lorimer,$^{8,9}$
A. Melatos$^{10}$
\\
$^{1}$Dept. of Physics and Astronomy, University of British Columbia, 6224 Agricultural Road, Vancouver, B.C., V6T 1Z1, Canada\\
$^{2}$Jodrell Bank Centre for Astrophysics, School of Physics and Astronomy, University of Manchester, Manchester, M13 9PL, UK\\
$^{3}$Max-Planck-Institut f\"ur Radioastronomie, Auf dem H\"ugel 69, D-53121 Bonn, Germany\\
$^{4}$ASTRON, The Netherlands Institute for Radio Astronomy, Postbus 2, 7990 AA, Dwingeloo, The Netherlands\\
$^{5}$Anton Pannekoek Institute for Astronomy, Univ. of Amsterdam, Science Park 904, 1098 XH Amsterdam, The Netherlands\\
$^{6}$Faculty of Education, University of British Columbia, 2125 Main Mall, Vancouver, BC V6T 1Z4, Canada\\
$^{7}$CSIRO Astronomy and Space Science, Australia Telescope National Facility, Box 76, Epping NSW 1710, Australia\\
$^{8}$Department of Physics and Astronomy, West Virginia University, PO Box 6315, Morgantown, WV 26506, USA\\
$^{9}$Center for Gravitational Waves and Cosmology, Chestnut Ridge Research Building, Morgantown, WV 26505, USA\\
$^{10}$School of Physics, University of Melbourne, Parkville VIC 3010, Australia
}

\date{Accepted XXX. Received YYY; in original form ZZZ}

\pubyear{2017}

\begin{document}
\label{firstpage}
\pagerange{\pageref{firstpage}--\pageref{lastpage}}
\maketitle

\begin{abstract}
The young pulsar PSR~B1828$-$11 has long been known to show correlated shape and spin-down changes with timescales of roughly 500 and 250 days, perhaps associated with large-scale magnetospheric switching.  Here we present multi-hour observations with the Parkes and Green Bank Telescopes at multiple phases across the $\sim$500-day cycle and show that the pulsar undergoes mode-changing between two stable, extreme profile states.  The fraction of time spent in each profile state naturally accounts for the observed overall "shape parameter" (defined to be 0 for wide profiles and 1 for narrow ones); this and the variable rate of the mode transitions are directly related to the spin-down changes.  We observe that the mode transition rate could plausibly function as an additional parameter governing the chaotic behaviour in this object which was proposed earlier by Seymour \& Lorimer.  Free precession is not needed to account for the variations.
\end{abstract}

\begin{keywords}
pulsars: general -- pulsars: individual: PSR~B1828$-$11 
\end{keywords}



\section{Introduction}

A prime characteristic of radio pulsars, and the property that makes them valuable tools for the precision study of physics ranging from the neutron-star interior to cosmological structure formation, is the stability of their rotation periods.  Pulsar timing studies enumerate every rotation of a neutron star across years or decades with no ambiguity; this leads directly to high-precision measurements of spin periods, astrometric parameters and binary parameters, if applicable.  Radio pulsars also generally have extremely stable pulse profiles at any given observing frequency.  This property is exploited to provide well-measured pulse arrival times by cross-correlation with a low-noise ``standard profile''.

\begin{figure*}
	\includegraphics[width=7in]{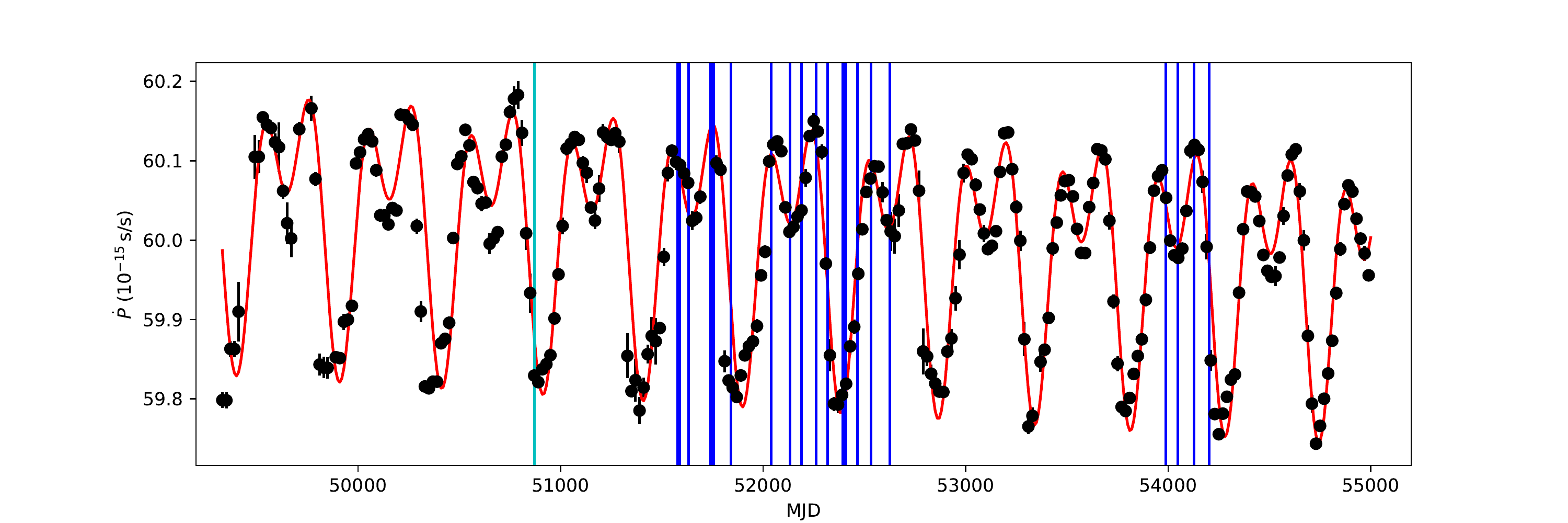}
    \caption{Black points with error bars show the pulse period derivative $\dot P$ for PSR~B1828$-$11, derived from the frequency-derivative values presented in \citet{lhk+10}.  Blue solid lines indicate the observing epochs for the Parkes and GBT data presented in this paper.  The red solid line is a fit to the $\dot P$ points incorporating two harmonically related sinusoids of decreasing period as well as a linear slope; see Section~\ref{sec:timing} and eq.~\ref{eq:sines}. The cyan solid line indicates the starting point of data included in the fit.}
    \label{fig:sampling}
\end{figure*}

There are some notable exceptions to the pulse stability rule, particularly when individual pulses are observed.  Some pulsars ``null," disappearing for anywhere from a few pulses to weeks at a time \citep[e.g.][]{bac70,klo+06}.  Another common behaviour is that of mode-changing, in which a pulsar displays two or more ``typical" profiles and switches between them with varying timescales \citep{bac70a}.  There is an indication of a relationship between these phenomena \citep[e.g.,][]{vkr+02} and also between the drifting subpulses that sum together to produce average profiles in a number of bright pulsars.

The young pulsar PSR~B1828$-$11 (PSR~J1830$-$1059) was discovered in a 20-cm survey of the Galactic plane conducted with the 76-m Lovell Telescope at Jodrell Bank Observatory in the late 1980s \citep{clj+92}.  Its pulse period is 405\,ms and its dispersion measure is close to $160$\, pc\,cm$^{-3}$.  PSR~B1828$-$11 has a characteristic age of 110\,kyr and an inferred surface magnetic field strength of $5\times 10^{12}$\,G, making it at first glance an unremarkable object among the pulsar population.  However, routine and eventually more intensive follow-up with the Lovell Telescope showed that it displays roughly periodic variations in its spin-down rate (period derivative), with periodicities of roughly 250 and 500 days.  These were correlated with the observed pulse shape, which ranged between ``wide" and ``narrow" extremes on the same timescales \citep{sls00}.  The initially favoured interpretation was that free precession was causing the spin axis to wobble, providing variable torque on the neutron star and allowing the observer to see different parts of the emission region over time.  This idea was not without problems, notably the expectation that pinned vortices in the superfluid internal to the neutron star would damp out such precession on timescales of several hundred precession cycles \citep[e.g.,][]{sha77,swc99}. Later authors \citep[e.g.][]{ja01,le01,lc02,lin06} argued that the existence of precession should instead be viewed as setting a limit on the amount of pinned superfluid in the star, with possibly large implications for neutron-star structure.  A complicating factor is that this pulsar underwent a glitch on 2009 July 29 \citep[MJD 55041; ][and \url{http://www.jb.man.ac.uk/~pulsar/glitches/gTable.html}]{elsk11}, which may force a requirement for pinned vortices after all \citep{jap17}.  

Meanwhile, other pulsars were beginning to show signs of correlation between profile shape and spin-down properties.  The most notable of these is PSR~B1931+24, which nulls for weeks on end, and moreover has a spin-down rate 50\% larger when the pulsar is on \citep{klo+06}. Further examples of this intermittent behaviour are now known in four other pulsars \citep{crc+12,llm+12,lsf+17}.  In the case of PSR~B1931+24, it seems clear that global changes in the magnetosphere affect both the torque on the neutron star and the ability of the pulsar to produce radio emission.  Using the Lovell Telescope pulsar data archive, \citet{lhk+10} argued that a similar type of magnetospheric switching must be taking place for six other pulsars including PSR~B1828$-$11.  This paper also presented evidence, from the average profiles of short observations, that PSR~B1828$-$11 likely spends most of its time in the extreme wide or narrow states, with successive days sometimes producing different profiles.  

Another model put forth to explain the phenomena seen in PSR~B1828$-$11 involves non-radial oscillations \citep{rc08,rmt11} in which pulsations of the neutron star with high spherical degree can modulate the emission and produce different pulse modes and/or drifting subpulses while simultaneously changing the torque.

More recent work has compared the Bayes factors for a precession model and a magnetospheric switching model for PSR~B1828$-$11, not taking into account any short-term changes and finding a weak preference for the precession model \citep{ajp16}.  \citet{ajp17} further show that the precession period is decreasing, which would imply an increase in the stellar deformation. \citet{jon12} argues that state switching and precession are fully compatible, with the latter driving changes in the time spent in each magnetospheric state. \citet{khjs16} similarly advocate that PSR~B1828$-$11 is undergoing precession which in turn is regulating the magnetospheric switching.

Here we report on multi-epoch long observations of PSR~B1828$-$11 with the Parkes and Robert~C.~Byrd Green Bank Telescopes which present evidence for mode-changing which differs from epoch to epoch, relate the mode-changing to the timing variability and show that the chaos model may be a good descriptor of the data.  Preliminary analyses of some of these data were presented in \citet{sakl03} and \citet{lyn13} and a brief overview of these results can be found in \citet{slk+18b}. In \S 2, we describe the observations and analysis. In \S 3, we present our results which are then discussed in detail in \S 4. Finally, in \S5, we draw conclusions and suggest directions for future work. 

\section{Observations and Data Reduction}

At the 64-m Parkes telescope, data were taken at 1390 MHz between 2000 Feb.\ 5 and 2002 Dec.\ 17 with a $2 \times 512 \times 0.5\,$MHz filterbank which provided 1-bit samples every 0.25\,ms.  The two polarizations were summed before digitization and the data are not flux-calibrated.  Most of these observations are about 4 hours in length, with the first ones broken up into several short files, while the data spans on a few days are shorter.                          
At the 100-m Green Bank Telescope (GBT), data were collected between 2006 Sept.\ 11 and 2007 April 12 with the Berkeley-Caltech Pulsar Machine \citep[BCPM; ][]{bdz+97} with 96 1-MHz frequency channels centred at 1400\,MHz.  Polarizations were summed in hardware and the data stream was sampled every 72\,$\mu s$. All these observations were about 4 hours in length.  Accompanying short calibration files were taken using a 25\,Hz switched diode with temperature 1.6\,K.  Uncalibrated data were used to determine the switching states and derived information, but flux-calibrated summed profiles and the resulting flux densities are presented later in this work.

The observation dates were chosen to sample the roughly 500-day period-derivative cycle.  In Fig.~\ref{fig:sampling} we plot the observing epochs over the timing and average-pulse-width data presented in \citet{lhk+10}.  Observations and basic properties are summarized in Table~\ref{tab:obs}, where we also include the phase of each observation relative to a fit to the $\sim 500$-day $\dot P$ cycle (see Section~\ref{sec:timing}) as computed for the integer MJDs.

\begin{table*}
	\centering
	\caption{Observations and mode-switching properties. Uncertainties in the last quoted digit(s) are given in parentheses. The entries in the ``Valid Data'' column are smaller than those in the ``Length'' column for those epochs in which the data are split into several short files.  Because {\tt dspsr} labels the integrations with integer multiples of 10 seconds, the data length values are generally very slight overestimates.}
	\label{tab:obs}
	\begin{tabular}{cccrrcrrr} 
		\hline
		MJD & Telescope &  $\dot P$ cycle & Length & Valid  & Average  & Number of & Transition & Peak ratio \\
		 &  & phase & (s) & data (s) & state & transitions & rate (s$^{-1}$) &  (W/N) \\
		\hline
51579 & Parkes & 0.330(5) &  1510 &   1510 & 0.901 &     14 &  0.00927 & 0.74 \\ 
51588 & Parkes & 0.348(5) &  4840 &   4520 & 0.781 &     83 &  0.01836 & 0.64 \\ 
51633 & Parkes & 0.440(5) & 11920 &  10610 & 0.445 &     45 &  0.00424 & 0.48 \\ 
51742 & Parkes & 0.662(5) &  3320 &   2980 & 1.000 &      0 &  0.00000 & 0.0 \\ 
51757 & Parkes & 0.693(5) & 10180 &   9050 & 1.000 &      0 &  0.00000 & 0.0 \\ 
51842 & Parkes & 0.867(5) & 14000 &  12000 & 0.021 &      2 &  0.00017 & 0.46 \\ 
52041 & Parkes & 0.275(5) &  6620 &   6010 & 1.000 &      0 &  0.00000 & 0.0 \\ 
52132 & Parkes & 0.462(5) & 13560 &  11140 & 0.458 &     36 &  0.00323 & 0.45 \\ 
52189 & Parkes & 0.580(5) & 15610 &  13180 & 0.351 &     10 &  0.00076 & 0.46 \\ 
52261 & Parkes & 0.728(5) & 14400 &  14400 & 1.000 &      0 &  0.00000 & 0.0 \\ 
52318 & Parkes & 0.846(5) & 13070 &  13070 & 0.000 &      0 &  0.00000 & --- \\ 
52394 & Parkes & 0.003(5) & 14060 &  14060 & 0.000 &      0 &  0.00000 & --- \\ 
52408 & Parkes & 0.033(5) & 13950 &  13950 & 0.000 &      0 &  0.00000 & --- \\ 
52466 & Parkes & 0.152(5) & 14400 &  14400 & 0.287 &      1 &  0.00007 & 0.46 \\ 
52532 & Parkes & 0.289(6) & 10610 &  10610 & 1.000 &      0 &  0.00000 & 0.0 \\ 
52625 & Parkes & 0.483(6) &  1350 &   1350 & 0.533 &      4 &  0.00296 & 0.48 \\ 
53989 & GBT & 0.358(10) & 15650 &  15650 & 0.473 &    152 &  0.00971 & 0.54 \\ 
54047 & GBT & 0.482(11) & 15900 &  15900 & 0.379 &     29 &  0.00182 & 0.46 \\ 
54127 & GBT & 0.653(11) & 15270 &  15270 & 1.000 &      0 &  0.00000 & 0.0 \\ 
54202 & GBT & 0.814(12) & 17050 &  17050 & 0.119 &     12 &  0.00070 & 0.45 \\ 
		\hline
	\end{tabular}
\end{table*}

\begin{figure}
	\includegraphics[width=\columnwidth]{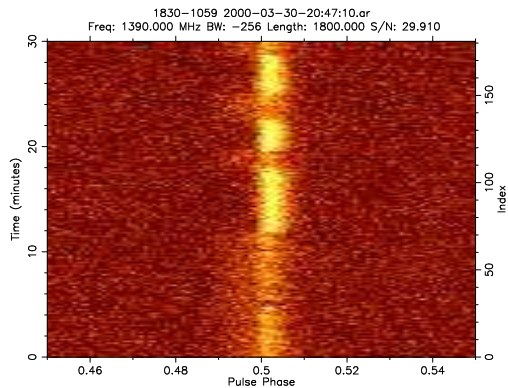}
    \caption{Intensity as a function of pulse phase and time/sub-integration for one of the Parkes observations on MJD 51633 (2000-Mar-30). The 10\% of pulse phase surrounding the pulse is displayed.  Six transitions between wide (fainter) and narrow (brighter) states can be readily identified.}
    \label{fig:pks}
\end{figure}

For each epoch, the filterbank search-mode data were folded using \verb+dspsr+ \citep{vb11} with 10~s sub-integrations and a dispersion measure of $159.7$\, pc\,cm$^{-3}$ \citep{sls00}.  This sub-integration length allowed the straightforward identification by eye of ``wide" and ``narrow" states within each observation using  PSRCHIVE's \verb+pav+ and \verb+paz+ programs \citep{vdo12}.  An example plot of a short Parkes data file is shown in Fig.~\ref{fig:pks}; the times at which the mode changes occur are just as clear in the other data sets.

\section{Results}
\subsection{Mode changes} \label{sec:modes}

\begin{figure*}
	\includegraphics[width=17cm]{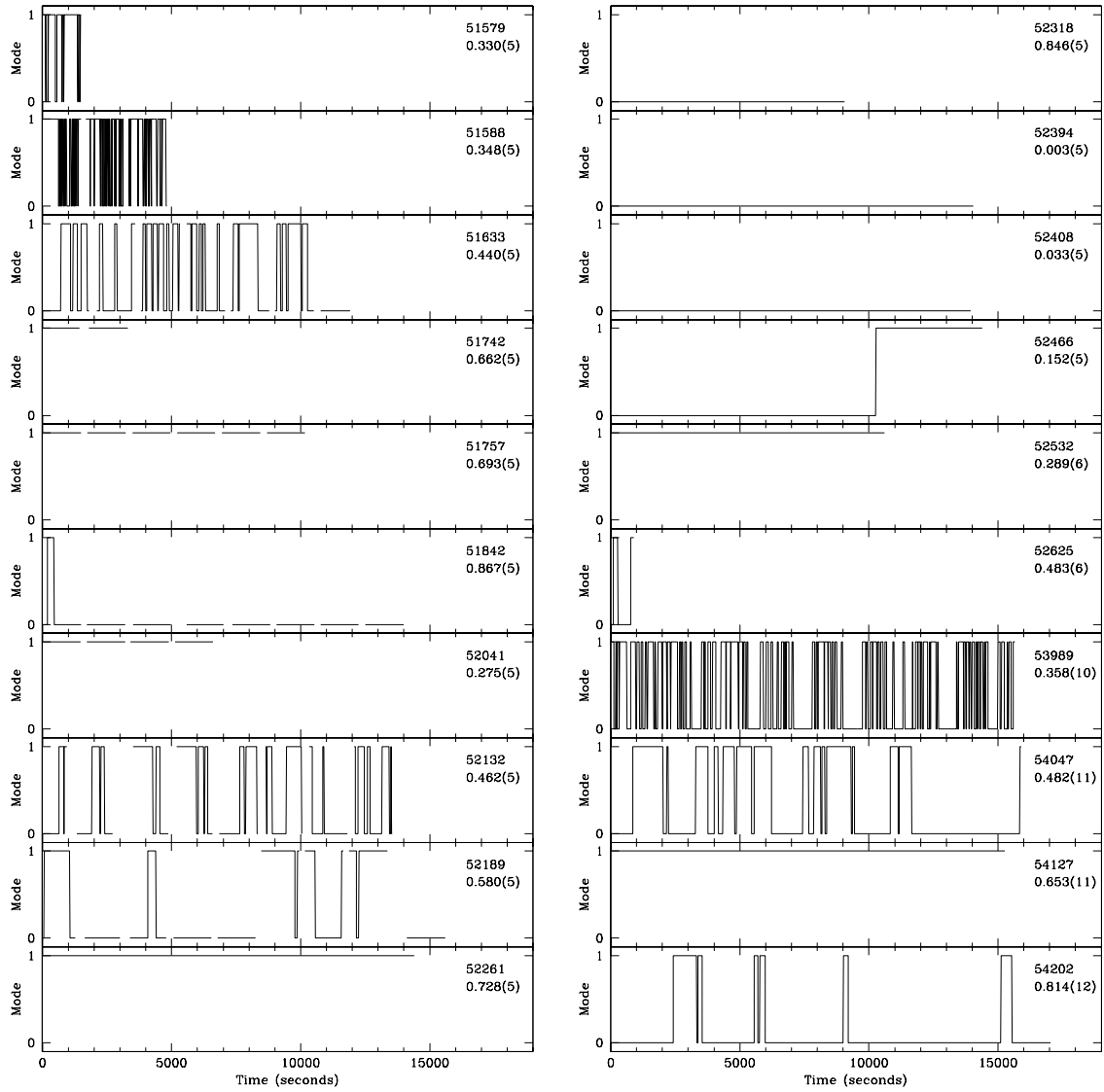}
    \caption{Diagram of observed modes (0 = wide, 1 = narrow) for each epoch, with MJD indicated to the left of each panel.  The first 16 panels (referring to column 1 and then column 2) are derived from Parkes data and the last 4 from GBT data.  There was significant radio frequency interference in the MJD 53989 observation at the GBT, but this caused uncertainty in the classification of only about 5 segments of data.  The observation MJD and the $\dot P$ cycle phase from Table~\ref{tab:obs} are shown in the top right-hand corner of each panel.}
    \label{fig:states}
\end{figure*}

In Fig.~\ref{fig:states} we show the switches between wide (0) and narrow (1) modes at each epoch.  Breaks in the plots occur when an observation is broken up into multiple data files.  It is clear that the fraction of time spent in each mode varies from epoch to epoch, along with the frequency of transitions.   Table~\ref{tab:obs} includes the average mode value along with the transition rate.

\begin{figure}
	\includegraphics[width=\columnwidth]{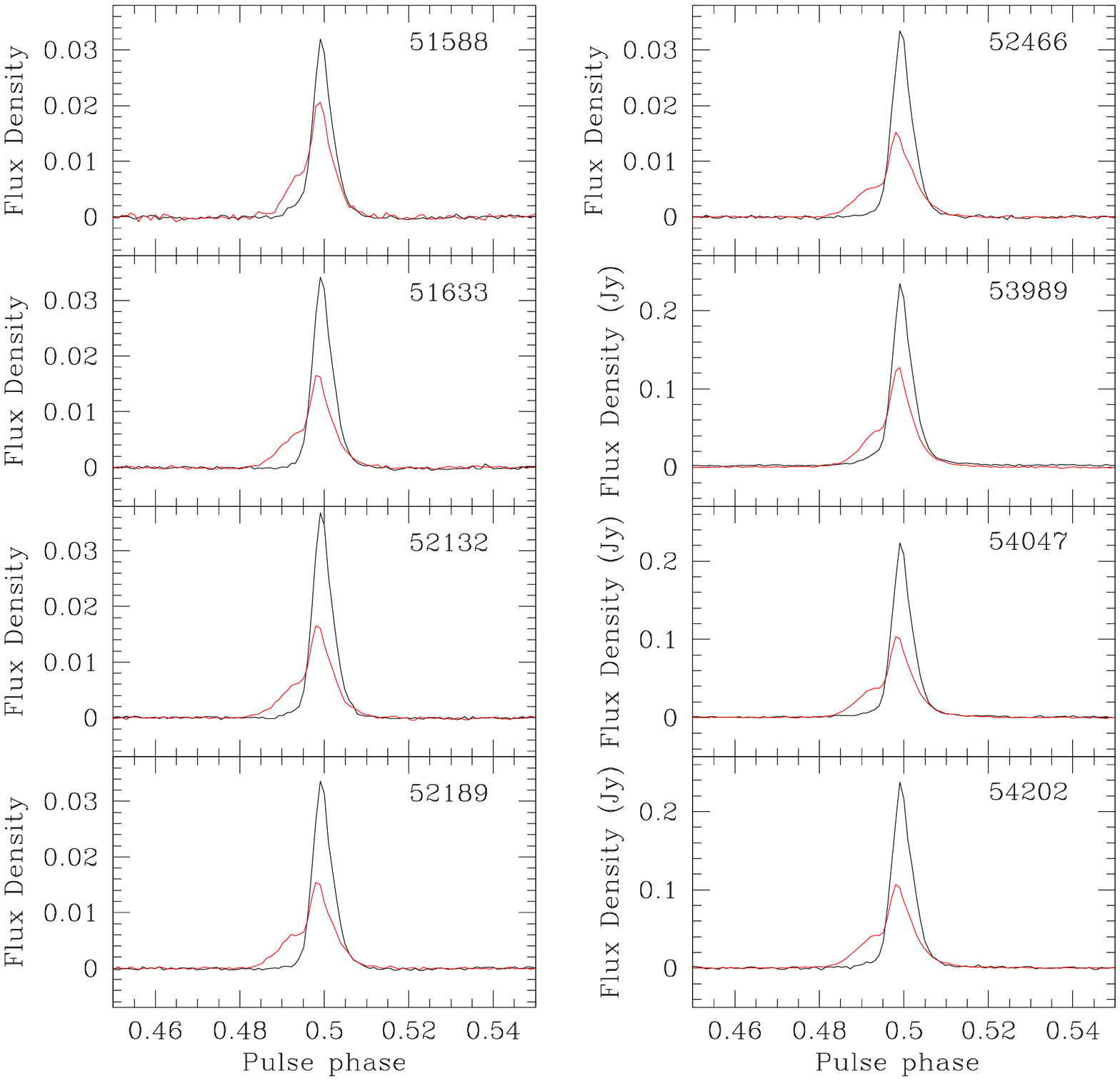}
    \caption{Cumulative narrow (black) and wide (red) profiles for each observing epoch with significant switching and long duration.  The observation MJD is shown in the top right-hand corner of each panel.  The first 5 panels are derived from Parkes data; the last 3 panels are from GBT data.  The Parkes profile flux densities are given in arbitrary units that are consistent from epoch to epoch.  The GBT data are flux-calibrated using a reference noise diode.}
    \label{fig:profs}
\end{figure}

At each epoch, the data files were subdivided into narrow and wide subfiles, and then summed to produce average profiles.  These are shown in Fig.~\ref{fig:profs}.  Although the Parkes data are not flux-calibrated (the levels are roughly consistent from epoch to epoch), it appears that the overall pulse heights are reasonably stable.   Table~\ref{tab:fluxes} presents the flux densities for the cumulative wide and narrow profiles for each GBT epoch.  The two days with long-duration sequences (see Fig.~\ref{fig:states}) have  similar flux density ratios between the wide and narrow pulses. The small overall difference in flux density may be due to refractive scintillation.  The epoch with the most rapid transitions, MJD~53989 (2006-Sept.-11), has a larger relative flux density in the wide profile; this is consistent with the higher peak of wide profile in Fig.~\ref{fig:profs}.  Note that a similarly higher-peaked wide profile occurs for the Parkes epoch with the most rapid transitions, namely MJD~51588 (2000-Feb.~14).  Whether or not this represents a true change in the wide-mode profile is not immediately clear, given the presence of significant radio-frequency interference on MJD~53989 and the inevitable inclusion of some narrow pulses within the wide-designated sub-integrations due to the 10-second sub-integration length.  This latter systematic would have the effect of increasing the height of the wide profile while simultaneously reducing the height of the narrow one.  The apparent small shape changes in the wide component may be due to a combination of noise and, again, the admixture of the narrow pulses contaminating some sub-integrations.   The small trailing ``shoulders'' on the narrow components on these two days also suggest such cross-contamination. In general, it is not possible to distinguish single pulses in these data (but see \S~\ref{sec:lrfs}), and in any case we are most interested in the pulse envelope.  The narrow and wide pulse profiles look very similar in shape at all epochs.

\begin{figure}
	\includegraphics[width=\columnwidth]{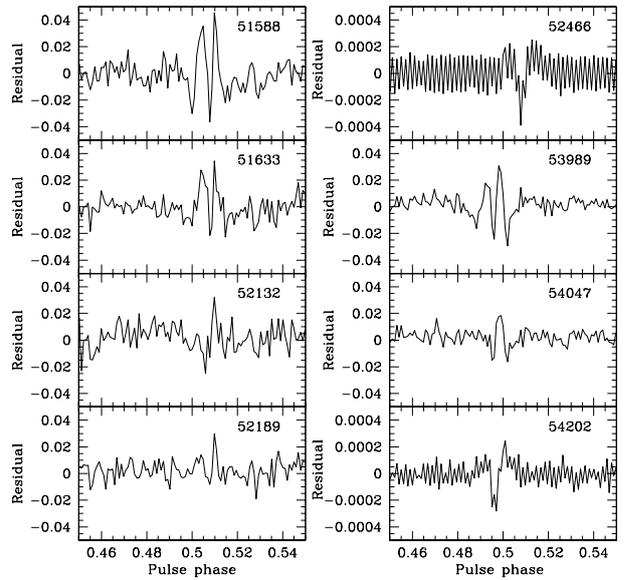}
    \caption{Difference profiles relative to a linear-combination fit to wide and narrow standard profiles for each observing epoch with significant switching and long duration.  The observation MJD is shown in the top right-hand corner of each panel.  The first 5 panels are derived from Parkes data and use the profiles from MJD 52466 for the standard profiles; the last 3 panels are from GBT data and use the profiles from MJD 54202 for the standard profiles.  The different choice of standard profile and hence fiducial point explains the slight offset in phase between the Parkes and GBT residuals; none of the plots are perfectly aligned with their counterparts in Fig.~\ref{fig:profs}. The vertical scale in each panel is the fraction of the peak of the cumulative profile at that epoch.}
    \label{fig:profres}
\end{figure}

To quantify this similarity, we have determined how well the per-epoch average profiles are fit by linear combinations of the extreme wide and narrow states, choosing the pairs of profiles on MJD 52466 for the Parkes epochs and on MJD 54204 for the GBT epochs as the standard profiles, as these dates had good representation of each state, with only a few, completely unambiguous changes.  The resulting profile residuals are shown in Fig.~\ref{fig:profres}. Their small relative amplitudes demonstrate that the two extreme profiles are sufficient to model the pulse.  This leads us to conclude that there are most likely only two profile envelopes visible to us no matter what the $\dot P$ phase.  This argues against a precession interpretation in which the impact angle to the magnetic pole is smoothly varying.

\begin{table}
	\centering
	\caption{Flux densities for the GBT data.  Uncertainties are estimated to be about 10\%.}
	\label{tab:fluxes}
	\begin{tabular}{ccccc} 
		\hline
		MJD & Wide & Narrow & Total & Ratio\\
		 & (mJy) &  (mJy) &  (mJy) & (W/N) \\
		\hline
		53989 &  1.3 &1.5  & 1.4  & 0.85 \\
        54047 & 1.0 & 1.3 & 1.13  & 0.74 \\
        54127 & --- & 1.5 & 1.5 & --- \\
        54202 & 1.07 & 1.5 & 1.12 & 0.73 \\
		\hline
	\end{tabular}
\end{table}

\begin{figure}
	\includegraphics[width=\columnwidth]{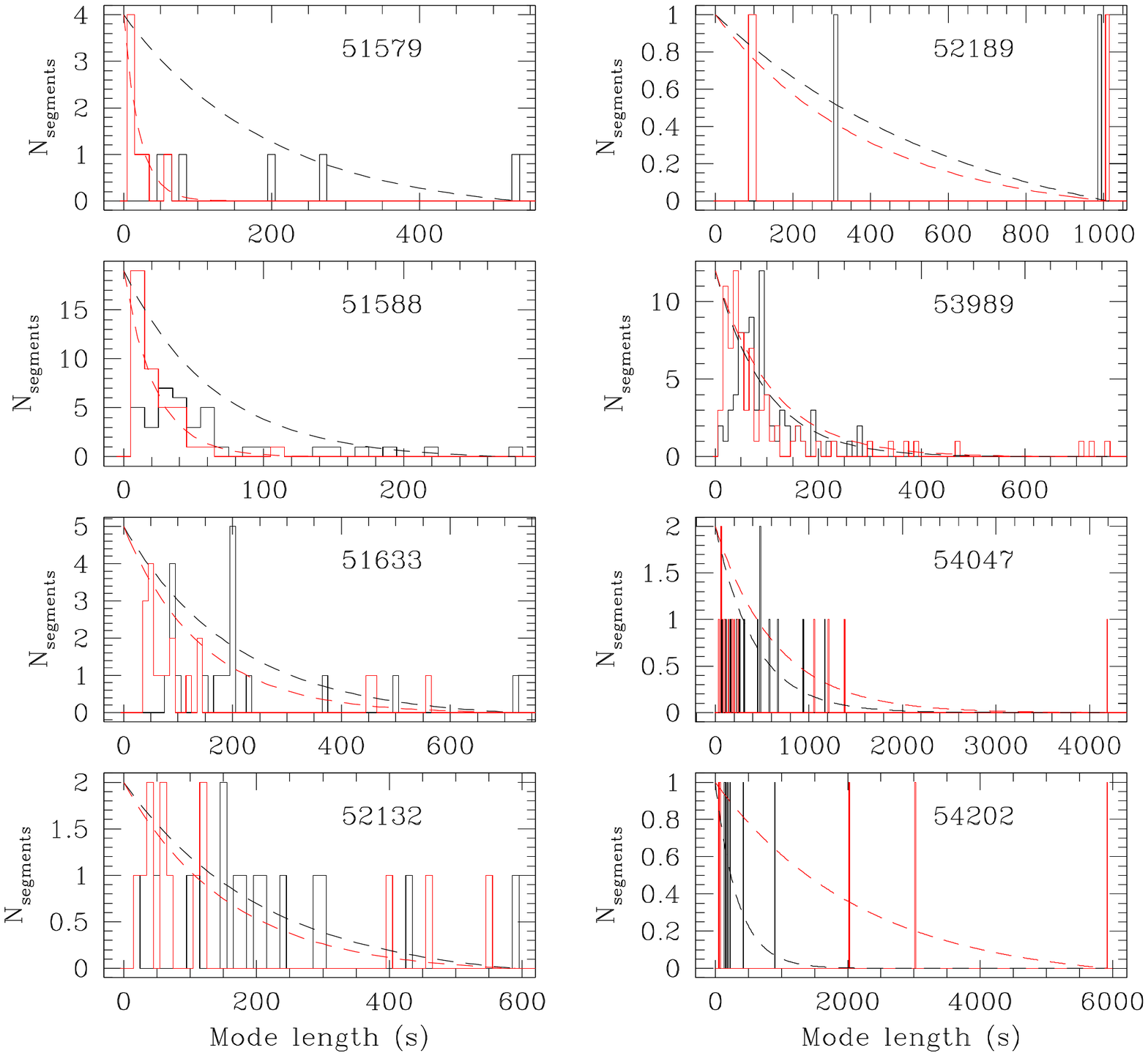}
    \caption{Histograms of the segment durations for the narrow (black) and wide (red) sequences of fully-observed sub-integrations for each observing epoch with significant switching. The observation MJD is shown in the top right-hand corner of each panel.  The first 5 panels are derived from Parkes data; the last 3 panels are from GBT data. The dashed lines show the state-change probabilities (with arbitrary vertical scaling) for 2-state Markov processes with means corresponding to the histogram means (see Eq.~1 of \citet{cor13}).}
    \label{fig:markov}
\end{figure}

Histograms of the wide and narrow segment durations derived from the fully-observed wide and narrow segments (i.e., those not at the boundaries of the observation) seen for each epoch with significant switching are shown in Fig.~\ref{fig:markov}. With these we plot the probability density functions (with arbitrary scaling) for two-state Markov processes with means derived from the histograms, as described in \citet{cor13}.  In three cases (the wide states on MJDs 51579, 51588 and 53989 -- all at roughly the same phase in the $\dot P$ cycle), these curves appear to be reasonable matches to the distributions, but on most days that is not the case.  The histograms do not appear to be well fit in general by exponential distributions, as would be expected if the mode transitions followed Poisson statistics.  There is often a pattern in which the wide sequences are more prone to very short durations, but this is not universally true. 

We carried out 2-sided Kolmogorov-Smirnov tests using the \verb+scipy+ routine \verb+ stats.ks_2samp+, to check whether the complete wide and narrow segments were drawn from different distributions, and whether the narrow and wide segment distributions differ from epoch to epoch. We summarize the results in Table~\ref{tab:KStest}. Overall there is evidence for some epochs having similar state durations and some having very different ones, as might be expected from the data in Fig.~\ref{fig:states}.  When considering these results, it should be borne in mind that MJD 51579 has very few full segments (7 wide segments and 6 narrow).

\begin{table*}
	\centering
	\caption{Evidence for different distributions for wide and narrow state segment durations.  Entries on the diagonal refer to the wide and narrow distributions on the relevant day.  Off-diagonal entries are for the paired wide/narrow distributions on those days.  ``Weak'' (W) evidence refers to $p \stackrel{<}{\sim} 0.01$ and ``strong'' (S) evidence to $p \stackrel{<}{\sim} 0.0001$ for different distributions.  Here, $p$ refers to the p-value returned from the KS test which refers to the probability that the two samples are consistent with being drawn from the same parent distribution. A dash indicates no evidence for difference ($p>0.01$)}.
	\label{tab:KStest}
	\begin{tabular}{ccccccccc} 
		\hline
		 MJD & 51579 & 51588 & 51633 & 52132 & 52189 & 53989 & 54047 & 54202 \\		 
		\hline
        51579 &  W	&	--/--	& W/--	 &	W/-- & 	W/-- & 	--/--	&	W/--&	W/--	\\
        51588 & 	&	W	&  S/S	 &	W/S	& 	W/-- & S/W	&  --/S &	W/W		\\
        51633 &  	&		&	W	 &	--/--	& 	--/--	 & --/W &	--/--	&	--/--		\\
        52132 & 	&		&		 &	--	& 	--/--	 & --/W	&	--/--	&	--/--		\\
        52189 &  	&		&		 &		& 	--	 & --	&	--/--	&	--/--		\\
		53989 &  	&		&		 &		& 		 & 	W	& --/S &	S/W		\\ 
        54047 &  	&		&		 &		& 		 & 		&	--	&	--/--		\\
        54202 &  	&		&		 &		& 		 & 		&		&	--		\\
		\hline
	\end{tabular}
\end{table*}

\begin{figure}
	\includegraphics[width=\columnwidth]{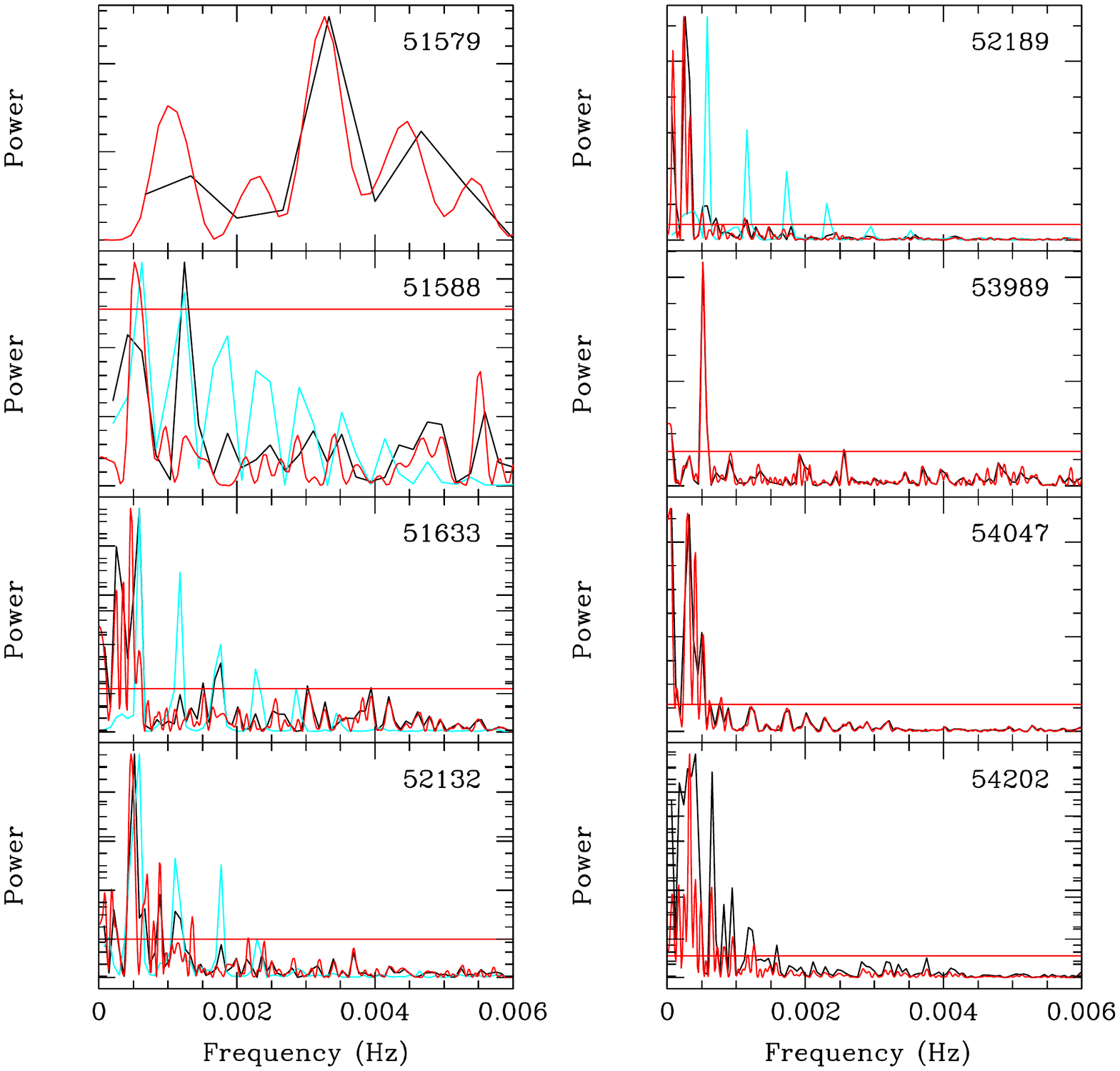}
    \caption{Power spectra (black curves) from the Fourier Transforms of the sequence of states for each observing epoch with significant switching. The observation MJD is shown in the top right-hand corner of each panel.  The cyan lines indicate the windowing function (arbitrary scale; DC component omitted) where the observations are broken into short scans.  The red curves show the Lomb-Scargle periodograms for the same datasets, with the red horizontal lines indicating estimates of false-alarm probabilities of 0.0001.  The first 5 panels are derived from Parkes data; the last 3 panels are from GBT data.}
    \label{fig:dftlsplot}
\end{figure}

In an effort to explore the long-period spectral content within the observations, we computed discrete Fourier Transforms of the state sequences for each epoch with rapid changes along with Lomb-Scargle periodograms using the \verb+astropy+ routine {\tt LombScargle} \citep{art+13}.  These are shown in Fig.~\ref{fig:dftlsplot}.  While the multiple short observations employed in the earliest Parkes data sets result in some unfortunate windowing (cyan curves), it is nevertheless clear that there are significant periodicities of the order of 30 minutes to an hour in several of the data sets.  In Table~\ref{tab:periodicities} we provide the frequencies and corresponding numbers of pulse periods for the most dominant well-defined peaks with false-alarm probabilities of less than 0.0001 in the Lomb-Scargle analyses, leaving out those which coincide with the relevant window functions. There must be some physical significance to the preference for state-change patterns lasting 30 to 60 minutes; this clearly demonstrates the need for routine long observations in order to capture and study this phenomenon properly.

\begin{table}
	\centering
	\caption{Dominant significant periodicities and corresponding timescales and numbers of pulse periods found in epochs with multiple transitions.}
	\label{tab:periodicities}
	\begin{tabular}{ccrr} 
		\hline
		MJD & Frequency & Timescale & Periods \\
            &      (Hz) &     (s)   & \\
		\hline
        51633  & 0.0002435 & 4108 & 10140 \\
          	   & 0.0003610 & 2770 & 6340 \\
               & 0.0004618 & 2165 & 5345 \\
        52132  & 0.0004649 & 2151 & 5311 \\
        52189 & 0.0000833 & 12005 & 29641 \\
              & 0.0002372 & 4216 & 10410 \\
		53989 & 0.0005179 & 1931 & 4768 \\
        54047 & 0.0002958 & 3381 & 8347  \\
              & 0.0004091 & 2444 & 6036 \\
              & 0.0005223 & 1915 & 4727 \\
        54202 & 0.0003228 & 3098 & 7649 \\
		\hline
	\end{tabular}
\end{table}

\subsection{Relationship to timing} \label{sec:timing}

\begin{figure}
	\includegraphics[width=\columnwidth]{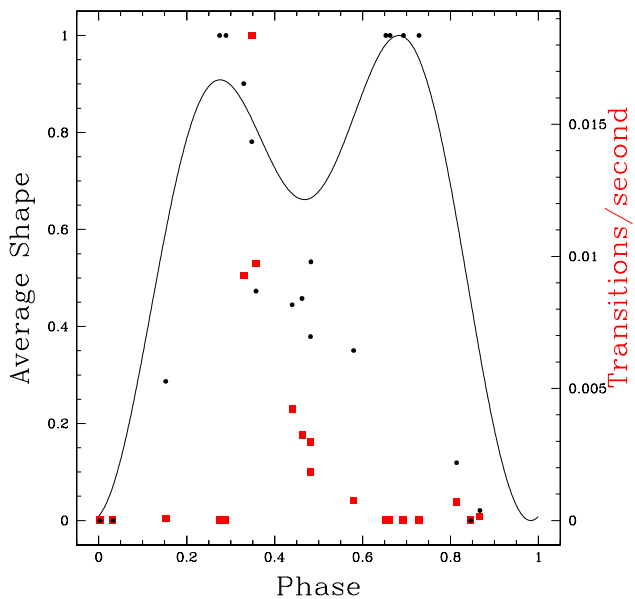}
    \caption{The average shape (black points; wide=0, narrow=1) and mode transition rate (red points) for each epoch as a function of the phase of the $\sim$500-day periodicity derived from the $\dot P$ fit described in Section~\ref{sec:timing} and represented by the black solid line.}
    \label{fig:avgratephase}
\end{figure}

The pulse period derivative for this pulsar has strong periodicities at roughly 500 and 250 days, as shown by \citet{sls00} and \citet{lhk+10}.  As shown by \cite{ajp16}, the periodicity decreases roughly linearly during the timeframe covered by the dataset. In order to plot our average shape values and transition rates against a consistently-defined phase, we first convert the \citet{lhk+10} frequency-derivative data to period derivatives $\dot P$, then carry out a least-squares fit with a linear slope (implying a period second derivative) as well as two harmonically-related sinusoids with linearly decreasing periods: 
\begin{eqnarray}
\dot P = A + B(t-52930) +C \cos\left(\frac{2 \pi (t-t_1)}{P_s+\dot{P_s}\times(t-52930)}\right) \nonumber \\+ D \cos\left(\frac{2 \pi (t-t_1)}{(P_s+\dot{P_s}\times(t-52930))/2}+\phi_2\right), \label{eq:sines}
\end{eqnarray}
\noindent where $P_s$ is the period of the cycle and $\dot P_s$ its time derivative, $t_1$ is a reference time, A, B, C, and D are constants, and $\phi_2$ is a phase offset.  The fit is shown overplotted on the $\dot P$ data in Fig.~\ref{fig:sampling}.  The reduced-$\chi^2$ of the weighted fit was 36.1, but we present the nominal uncertainties on the fit parameters in Table~\ref{tab:timingsines} as it is clear that the phases of the sinusoids are well-captured by this fit.  We used the MJD range MJD 50870--54990 in the fit. In Fig~\ref{fig:sampling} the fit is shown extended to the earlier points in the data set; if these earlier points are included in the fit, the reduced-$\chi^2$ is somewhat worse at 40.8, therefore we restrict the fit to the later part of the data set. For each observing epoch, we compute the phase relative to the $\sim 500$-day $\dot P$ cycle, taking into account the cycle period decrease.  Note that phase 0 corresponds approximately to the times at which the profile is purely in the wide state.   In Fig.~\ref{fig:avgratephase} we plot the average shape and the transition rate as a function of this phase.  It is clear that the pattern of the average shape is a very good match to that of the period derivative.  When the profile is wide, at phase 0, the spindown rate is lowest. At later phases (near 0.3 and 0.75) in the $\dot P$ cycle, when the profile is narrow, the pulsar spins down faster.  This reinforces the findings of \citet{sls00} and \citet{lhk+10}. The arguments presented in Sec.~\ref{sec:modes} show conclusively that the varying widths reported in these papers were in fact due to changing fractions of time spent in each of the two emission modes.  This represents a profound connection between the short- and long-term behaviours in this pulsar. Meanwhile, the rate at which transitions occur within an observation also has a $\sim$500-day periodicity, with small amounts of activity at most phases but a peak in activity around phase 0.35.  While the second half of the $\dot P$ cycle is not quite as well-sampled, the transition rates that we do measure are generally lower.  We discuss the implications of this difference in Section~\ref{sec:chaos}.

\begin{table}
	\centering
	\caption{Fit to period derivative data.  Uncertainties in the last quoted digit are given in parentheses.}
	\label{tab:timingsines}
	\begin{tabular}{cc} 
		\hline
		Parameter & Value \\
		\hline
		A  & 59.993(2) $\times 10^{-15}$\,s/s  \\
        B  & $-1.6$(2)  $\times 10^{-15}$\,s$^{-1}$  \\
        C & $-0.118$(3) $\times 10^{-15}$\,s/s  \\
        D & $-0.102$(3) $\times 10^{-15}$\,s/s  \\
        $P_s$ & 482.9(9) days\\
        $\dot P_s$ & $-0.0048$(5) days/day \\
        $t_1$ & MJD 51907(2) \\
        $\phi_2$ & 0.29(5) rad \\
		\hline
	\end{tabular}
\end{table}

\section{Discussion}

\subsection{Compatibility with precession}\label{sec:precession}

The mechanism underlying the long-term spin-down and average-profile changes in PSR~B1828$-$11 is still in dispute in the literature.  Free precession remains a popular hypothesis.  This process involves a misalignment of the spin and angular momentum axes of the neutron star, causing the observer to have different impact angles to the spin and magnetic axes as a function of time, with the angle between these two axes remaining constant.  This can naturally explain a smooth variation in pulse shape as different parts of the emission region come into view.  There have been numerous attempts to model the various angles and emission-beam geometries, starting with \citet{le01}, who advocated for an hourglass-shaped beam as also inferred for the geodetically precessing Hulse-Taylor binary pulsar \citep{wt02}.  In fact, variations on an hour-glass beam shape appear to be common to most (biaxial) precession models for this pulsar \citep{ajp16,ajp17}, though we note that \citet{alw06} derived a triaxial model with a core/cone beam shape. Intuitively, the hourglass shaped is required because the pulsar would pass from a ``completely wide state'' through a ``completely narrow" one to a ``slightly wide'' one and back through the ``completely narrow'' state as it goes through a $\dot P$ cycle.  

These previous beam-shape models made use of either ``shape parameters'' \citep{sls00} averaged over a certain range of the $\dot P$ cycle, or else the 10\% width of the average pulse profile as determined by either a months-long average or else individual observations \citep{lhk+10}.  Both these quantities take on a continuum of values and are subject to uncertainties in the shape/width modeling for any given profile.  It is therefore not surprising that a model which incorporates smooth changes in the pulse profile (that is, precession) appears in these analyses to be favoured over models allowing rapid switching between binary states \citep{ajp16}.

In contrast, we have demonstrated in Section~\ref{sec:modes} that: (1) there are abrupt, clean transitions between states in all the observations where both states were observed (modulo slight cross-contamination in the 10-s integrations on days where the switching is extremely rapid); (2) all the observations which include both states can be modeled satisfactorily as a linear combination of the two extreme states as determined from epochs with very few and completely unambiguous switches.  These two properties imply that there is no difference in the overall pulse envelope from epoch to epoch.  This result is completely compatible with a non-precessing pulsar for which we always see the same part of the emission region and for which the mode-switching timescales vary across the $\dot P$ cycle.  We cannot rule out precession, but it would likely require a very small change in impact parameter in order to maintain the similar beam shape, and this fine-tuning combined with  Occam's razor leads us to strongly disfavour the precession hypothesis.  It would be interesting to see the analyses of \citet{ajp16,ajp17} redone incorporating the quantized nature of the profile switches.

\subsection{Similarly switching pulsars}\label{sec:similar}

As shown in \citet{lhk+10}, there are now numerous pulsars which appear to undergo similar forms of sometimes abrupt switching between profile and related spin-down states.  Notably, PSR~B1822$-$09 also shows evidence of rapid short-term switching \cite{lhk+10} and it is in fact known to undergo mode changes \citep[e.g.,][]{hkh+17}.  The latter study considered 8 lengthy observing sessions showing mode changes, and did not note any differences in mode durations from epoch to epoch.  The authors note, however, that all of their epochs were at times when the pulsar's spin-down rate was similar.  Other published observations of mode changing in PSR~B1822$-$09 \citep[e.g.,][]{bmr10,lmr12} indicate that the modes do sometimes last longer than the historical averages \citep{fwm81}; unfortunately the observing epochs appear to be outside the timing variation study of \citet{lhk+10}.  Further long observations, triggered by changes in $|\dot \nu|$, could prove very informative.

Two further pulsars which have now been extensively studied are PSRs~B0919+06 \citep{psw+15} and B1859+07 \citep{psw+16}.  Both pulsars show a ``flare'' profile state \citep{rrw06} which may be partly related to observed changes in spin-down in PSR~B0919+06 but apparently not in PSR~B1859+07.  \cite{worw16} go so far as to attribute the flare/``swoosh'' states to effects from ultra-dense binary companions.  Nevertheless, the average pulse profile state does have a relationship to the spin-down in each case. \citet{psw+15} proposed a 4-phase model for the spin-down variations in PSR~B0919+06, simulating pulse arrival times and demonstrating that the frequency derivative, which necessarily must be measured across a finite span of days, can be reproduced from such abrupt changes.  \cite{ajp16} use a similar multi-phase model when attempting to model the spin-down changes in PSR~B1828$-$11 and obtained a reasonable description of the window-averaged spin-down measurements.

Whether such long-duration spin-down states with abrupt changes accurately represent the physics of this pulsar is an important question.  Clearly, long-duration states with instantaneous changes in spin-down correctly describe what is seen in the long-term nulling pulsar PSR~B1931+24, although there are changes in the durations of the states \citep{klo+06,ysl+13}.  However, the mode-changing data that we present in this work clearly show: (1) changes in the average shape value over time; (2) changes in the rate of mode-changing over time; (3) a relationship between both sets of changes and the phase of the $\dot P$ cycle.  It therefore seems more plausible to infer that there are {\it short-term} changes in $\dot P$ corresponding to the mode changes.  As $\dot P$ can only be measured over an observing window of weeks, its variations are smoothed out to produce the shapes seen in Fig.~\ref{fig:sampling} (see \citet{ssw18} for an investigation of this effect).  In contrast, the shape and transition rate are sampled at particular phases \citep[this work]{lhk+10}.  Whether the changes in shape and transition rate are correlated at any given time -- that is, tracking the smoothed $\dot P$ changes -- or have a significant random element will have to be investigated with high-cadence observations with long tracks. 

\subsection{Possible chaotic behaviour}\label{sec:chaos}

\begin{figure}
 	\includegraphics[width=\columnwidth]{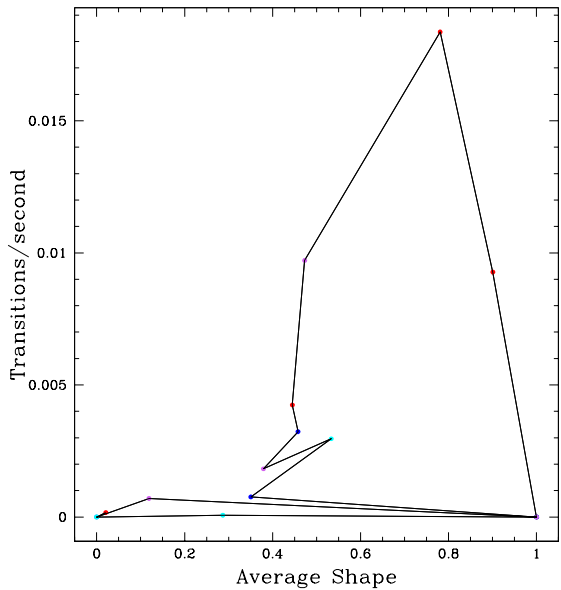}
    \caption{The mode transition rate as a function of the average shape (wide=0, narrow=1), with the points connected in order of occurrence during the $\sim$500-day $\dot P$ cycle. Different cycles are indicated by different-coloured points. Note that the (0,0) and (1,0) points consist of multiple points stacked together.  Allowing for some cycle-to-cycle variability, it appears that the $\dot P$-cycle trajectory consists of two loops on this plot.  Note that the three points with the largest transition-rate values occur at roughly the same $\dot P$-cycle phase, with two points in cycle 0 (red dots) and one in cycle 5 (purple dot).  Different choices of data length included in the Eq.~\ref{eq:sines} fit can change the relative ordering of these dots.  Given the expectation of some cycle-to-cycle variability in chaotic behaviour, we do not consider this a problem for our argument, but we choose to present this version of the fit and plot for maximum clarity.}
    \label{fig:avgrate}
\end{figure}

In Fig.~\ref{fig:avgrate} we plot the transition rate and shape data from Fig.~\ref{fig:avgratephase} against each other, with the solid line connecting points with increasing $\dot P$-cycle phase.  While the second half of the cycle is less well sampled, it appears that there are separate trajectories in this plane for the two halves of the $\dot P$ cycle, likely with some cycle-to-cycle variability.  Despite this lack of coverage, it is inescapable that the point (1,0) (narrow profile, zero transitions) is reached twice in the cycle.  This type of two-piece oscillation could potentially be explained by a set of nonlinear limit-cycle equations similar to the classic predator-prey equations \citep{lot20,vol26}.  Given the broader context of the identification of chaotic behaviour in the frequency derivative of PSR~B1828$-$11 \citep{sl13}, we note that the shape of the plot in Fig.~\ref{fig:avgrate} is also reminiscent of a projected chaotic attractor, and consider whether the characteristics of the pulsar can all fit with this description.

\citet{ajp17} suggest that the decrease in the~500-day periodicity can be attributed to changing deformation in a precessing pulsar.  However, we propose that the decrease can be accommodated within the chaotic behaviour.  First, as discussed in Sec.~\ref{sec:timing}, a linear decrease in cycle period is a poorer assumption over the full \citet{lhk+10} data set than over the segment starting at MJD 50890.  Furthermore, longer-term data will be needed to see if this trend continues.  We note that PSR~B0919+06 shows a decrease in its $\dot P$ cycle period as well \citep{psw+15}, with a reasonably abrupt change and pattern phase shift that are not easily explained.

\citet{sl13} argue that there must be three variables governing the time evolution of PSR~B1828$-$11.  They identify the spin-down rate as one of these variables, and suggest quantities associated with the magnetic field/pulse profile for a second and the internal superfluid for the third.  The average profile shape is clearly a proxy for the spin-down rate, as they are highly correlated.  Here we have qualitative evidence for the transition rate between modes (which is likely related to the current and/or magnetic field properties) as a second governing variable.  High-cadence data, preferably over multiple $\dot P$ cycles, will again be needed to explore this possibility in detail.

The two extreme profiles are well within the norm of pulsar profiles, and any single observation would simply prompt an observer to label this pulsar a mode-changer.  Our observations make clear that the rate of mode changing is time-variable in a nearly repeatable fashion.  Important questions are: what governs the transition rate and the average shape at any $\dot P$ cycle phase?  If the narrow and wide extremes are both metastable states, why are there two possible ``decay'' trajectories from the narrow state (to average shape ~0.5 and to average shape ~0, with very different transition rates), and how does the system know to alternate between these two trajectories?   We note that the two phases at which the profile appears purely narrow correspond to slightly different (short-term-averaged) values of $\dot P$. It seems inescapable that there must be yet another, as yet unknown, variable driving the system, as expected in the chaos description.  Given the glitch in 2009, it seems plausible that, as \cite{sl13} proposed, the internal structure of the neutron star may be involved.  The identification of a relationship between glitch activity and profile changes in PSR~J0742$-$2822 \citep{ksj13} is intriguing in this respect, although the connection may not be robust (Shaw et al., in prep).

\subsection{Physical mechanisms}\label{physics}

\citet{jon12} argues that the abrupt switching is due to changes in pair production, when the particle energy is pushed over a critical threshold (possibly a combination of the pair-production threshold and the work function for the neutron-star surface).  This would be compatible with changes seen in ordinary moding pulsars \citep[e.g.,][]{wmj07}.  \citet{jon12} further suggests that precession could cause the magnetosphere to prefer one switched state over another on the precession timescale. One possible corollary of this bias is the type of oscillation seen in PSR~B1828$-$11. While we cannot definitively argue against this mechanism, we have made the point above that precession appears unlikely to be a factor for this pulsar, given the lack of evidence for continuous pulse envelope changes.

It is difficult to interpret our results concretely in terms of current theories of pulsar electrodynamics. Substantial pulse profile changes do not necessarily imply substantial changes in the global magnetic geometry. Even if the radio beam is monolithic, relatively small magnetic distortions can steer large parts of it away from the line of sight. Equally, the radio beam may be composed of multiple, overlapping sub-beams, corresponding to emission regions at various latitudes, longitudes, and altitudes in the magnetosphere \citep{mr17}. Relatively small changes in what field lines carry the radio-emitting conduction currents can switch some of these regions on and off, producing pulse profile changes without global magnetic changes.\footnote{This is especially true, if the emission emanates from well within the light cylinder. Near the light cylinder, the magnetic geometry is significantly modified by the \citet{gj69} current along open field lines (and the displacement current), and it is harder to decouple current fluctuations from magnetic fluctuations.\label{footnote:lcyl}} Substantial changes in the spin-down torque can also be explained without substantial changes in the global magnetic geometry. The lever arm of the torque is the light cylinder. It is therefore possible to modify the torque by redirecting conduction currents in the vicinity of the light cylinder without changing their paths very much at lower altitudes. This scenario comes with the attractive feature that it is intrinsically nonlinear, consistent with the dynamics inferred from the data, e.g., Fig.~\ref{fig:avgrate}; current changes near the light cylinder induce magnetic changes locally (see footnote~\ref{footnote:lcyl}), which redirect the currents in turn, allowing naturally for feedback, instabilities, and multi-stable states.\footnote{Of course, just because such dynamics are allowed does not mean they occur; a specific, concrete feedback loop must first be identified self-consistently.} A nonlinear version of the circuit models developed by \citet{shi91} may prove to be a profitable framework within which to analyse the possibilities. The main challenge for models of this sort is that the dynamical time-scale of the magnetosphere is the spin period, even when nonlinear effects are included. Unless some (e.g., diode-like) circuit element can be argued to function on a slow timescale, perhaps set by anomalous resistivity as in planetary magnetospheres, it is hard to get quasiperiodic changes and system memory on the timescale of years. 

Long-term memory is easier to understand if it is modulated by processes in the crust, which push around the magnetic footpoints (and hence alter the global magnetic geometry), or channel conduction currents along different subsets of field lines, as discussed above. Both these scenarios would naturally lead to changes in the spin-down torque. The problem with crust-based scenarios is there is no obvious reason that they should be quasiperiodic \citep[c.f., ][]{hyt16}. It is easy to imagine slow, stochastic drifts, analogous to polar magnetic wandering on the Earth \citep{mac74,wh15}. But to get quasiperiodic dynamics, e.g. via a nonlinear limit cycle, one needs a feedback loop: the magnetospheric currents must modify the crust somehow. This is not easy to achieve, because the Lorentz forces are small compared to the elastic and hydrostatic forces that control the crust's structure. Dynamo action is a distant possibility, but the consensus is that it is quenched by stratification \citep[][but c.f.\ \cite {bra06}]{fmz15}. 

Without trying to determine what this underlying change agent is, \citet{tim10} proposed that relations between torque and profile may provide insights into the magnetospheric switching. Here, profile shape changes/nulling can have corresponding torque changes in the force-free magnetosphere model of pulsar emission, assuming that the magnetosphere has two or more quasi-stable states, with different closed-field regions and/or different current distributions in the open-field-line zone.  An intrinsic part of that model is, however, that a large zone implies high current and energy loss. For circular beams, a larger zone implies a wider profile. Thus, according to this model, wider-profile modes spin down faster than narrow profiles. If, in pulsars like the intermittent PSR B1931+24, the narrow mode misses Earth, the spin-down during this apparent null should be less than when the pulsar is visible. That is indeed the behavior that is observed in that pulsar. In PSR~B1828$-$11, however, we do see both modes. Contrary to what the \citet{tim10} model would predict, we find the narrow profiles show the highest spin-down. The high spin-down should occur in the mode with the larger open-field-line zone. Only if the beam in the narrow mode is more elongated than in the wide mode \citep[as seen in e.g. the fan beam producing the main pulse of PSR J1906+0746;][]{dkc+13} can this larger area be realized.

In summary, therefore, there is no single scenario which addresses every serious issue adequately: nonlinear limit cycles seated in the magnetospheric suffer from a short-time-scale problem, but longer-term crustal processes are unlikely to be cyclic. The best hope for untangling the problem remains better quality polarization data, which directly probe the local (as opposed to global) magnetic geometry in the radio-emitting region.

The non-radial oscillation model \citep{rmt11} may also be compatible with switching. In the case of PSR~B1828$-$11, changes in the oscillation quantum number $\ell$ would provide the changes in pulse shape.   Whether any of these emission models can be subject to chaotic behaviour is a subject for further theoretical investigation.  Identification of additional, unobserved governing variables and subsequent full modeling of the system remains a long-term goal of these studies.  

\subsection{Tests for periodic magnetospheric variations}\label{sec:lrfs}

In the preferred model above, the fast profile switching and the slower $\sim$500-day $\dot P$ periodicity are caused by magnetospheric changes. Beyond the mode changing described above, a number of further phenomena can be used to derive the physical properties of pulsar magnetic fields. Foremost in these are measurements of subpulse drift, and the periodicity at which these drift patterns recur. Changes in drift rates are especially insightful. In \citet{vt12}, for example, these variations were used to determine relative changes in the accelerating potential drop of only 10$^{-3}$. 

We investigated whether these magnetospheric characteristics could also change over our $\sim$500-day cycle. If we could detect, in one or both of the modes, any subpulse drift or other single-pulse periodicity, then these could be tracked over the 500-day cycle, and allow for mapping of the magnetosphere and open-field-line region. We first looked if any such signals are visible in the longest, brightest sets. For these we formed uncalibrated single-pulse time series from the GBT data taken with BCPM, and we identified MJD 54127 for the narrow mode, and MJD 54202 for the wide mode, as the brightest series free of radio interference.

\begin{figure}
        \centerline{
        \includegraphics[width=\columnwidth]{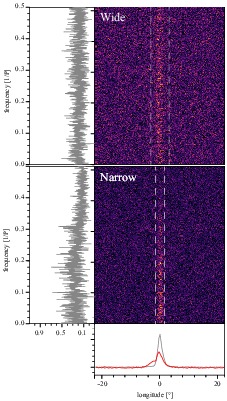}}
        \caption{The LRFS plots for the wide (top) and narrow (bottom) modes. Average profiles for both modes shown at the bottom panel. The main, color panels show the intensity of any modulations frequency as a function of pulse longitude. No features stand out. For the narrow mode observation, some low level RFI is visible over all phases for frequency 0.31 P$^{-1}$. The subpulse modulations frequencies in the main pulse, between the dashed lines, are collapsed and shown in the left panels, in units of 1/(pulse period). No significant peaks are detected.}
    \label{fig:lrfs}
\end{figure}

In each, we isolated bright consecutive  single pulses which we transformed to Longitude Resolved Fluctuation Spectra \citep[LRFS; ][]{bac70b}.  These Fourier transforms of consecutive pulses along each longitude can show which part (if any) of the pulse profile shows periodic behaviour such as drifting or other modulation.  We took series of 2048 pulses (marked red in Fig.~\ref{fig:states}) to maximise signal-to-noise ratio while limiting the wash-out of any periodicities over longer periods.  We applied the LRFS transformation using the \verb+Single Pulse Analysis+ toolkit \citep{sv17}, and the results are shown in  Fig.~\ref{fig:lrfs}. Compared to the large similarly derived sample in  \citet{wes06}, the results lack the features expected for periodic modulation. 

Given this non-detection in both modes, the prospects of using single-pulse modulation for further quantification of the $\sim$500-day magnetospheric changes are slim.

\section{Conclusions}

We have performed an exhaustive analysis of the variations in mode-changing seen at different observing epochs for PSR~B1828$-$11.  The overall pulse envelope is well-modeled by a linear combination of extreme wide and narrow profiles for each phase of the well-known $\sim$500-day $\dot P$ cycle.  This argues against precession, which would provide us with a different view of the emission region at different $\dot P$-cycle phases.  Histograms of state durations do not match particularly well to exponential, Poisson or Markov distributions.  While there do not appear to be strong patterns within individual observations, there is some evidence for a timescale of roughly 30 to 60 minutes in data from various epochs.  
We compute the average mode-switching transition rate for each epoch.  When plotted against the average shape value and viewed in order of increasing $\dot P$-cycle phase, the pattern resembles a projected double-loop chaotic attractor, matching the predictions of \citet{sl13} who identified chaotic behaviour in the spin-down value alone.  We therefore believe we have identified this transition rate, related to the state of the magnetosphere, as a second variable in the chaotic system.  We join \citet{sl13} in speculating that the neutron-star interior may provide the third governing variable. 

Confirmation of the transition rate as the second governing variable will require considerably more data than are presented here.  \citet{lhk+10} noted that PSR~B1828$-$11 frequently displayed opposite-extreme profiles in nearby observations.  Our data, with observations typically lasting 4 hours, show that half-hour observations are generally inadequate to capture the true average profile (and there may well be even more transitions with timescales longer than 4 hours).  Data stored in the usual pulsar timing setup, with the pulses integrated for minutes at a time, will not allow the identification of all the transitions in an observation.  What is needed is multi-hour time-series observations every day or two with a telescope with the sensitivity of Parkes or the Lovell Telescope, over at least one full $\dot P$ cycle.  Such dedicated observing time is of course hard to come by.  The upcoming CHIME telescope \citep{baa+14} will likely observe this pulsar on a daily basis, but it is a transit instrument operating at a frequency lower than has historically been used for this pulsar.  The best hope may be to use a subarray of the Phase I SKA telescope \citep[e.g.,][]{kja+15}. 

The possibility of chaotic behaviour in the mode transition rate and timing for PSR~B1828$-$11 opens up the question of this pulsar's place in the broader population.  \citet{sl13} found patterns suggestive of but not confirming chaotic behaviour in 3 other pulsars: PSRs B1540$-$06, B1642$-$03 and B1826$-$17, namely the best-sampled pulsars presented in \citet{lhk+10}.  Previous attempts to discern chaotic behaviour in pulsars, whether in timing behaviour \citep{hsc90} or in emission properties \citep{dw99} did not find evidence in favour of chaos, but longer and better-sampled data sets may make a difference. As discussed above, PSR B1822$-$09 deserves further study to investigate differences in mode-duration lengths.  Questions to be considered include whether the existence and timescales of these phenomena are related to the pulsar parameters such as period, age and magnetic field.  In-depth studies of these objects could prove very rewarding.

\section*{Acknowledgements}

IHS is supported by an NSERC Discovery Grant and by the Canadian Institute for Advanced Research.  Pulsar research at the Jodrell Bank Centre for Astrophysics and the observations using the Lovell Telescope are supported by a consolidated grant from the STFC in the UK. JvL received funding for this research from the Netherlands Organisation for Scientific Research (NWO) under project "CleanMachine" (614.001.301).  DRL was supported by NSF RII Track I award number 1458952.  The Green Bank Observatory is a facility of the National Science Foundation operated under cooperative agreement by Associated Universities, Inc.   The Parkes radio telescope is part of the Australia Telescope National Facility which is funded by the Australian Government for operation as a National Facility managed by CSIRO.  This research made use of Astropy, a community-developed core Python package for Astronomy (Astropy Collaboration, 2013).  We thank Ryan Hyslop, Jennifer Riley, Raymond Lum, Cindy Tam and Sarah (Traine) Keith for their work on earlier versions of the analysis.












\bsp	
\label{lastpage}
\end{document}